# Pressure-induced quantum phase transitions in topological insulator YbB$_6$

Yazhou Zhou[1], Dae-Jeong Kim[2], Priscila Ferrari Silveira Rosa[2], Qi Wu[1], Jing Guo[1], Shan Zhang[1], Zhe Wang[1], Defen Kang[1], Wei Yi[1], Chao Zhang[1], Yanchun Li[3], Xiaodong Li[3], Jing Liu[3], Peiquan Duan[4], Ming Zi[4], Xiangjun Wei[4], Zheng Jiang[4], Yuying Huang[4], Yi-feng Yang[1,5], Zachary Fisk[2], Liling Sun[1,5]†  & Zhongxian Zhao[1,5]†

[1]*Institute of Physics and Beijing National Laboratory for Condensed Matter Physics, Chinese Academy of Sciences, Beijing 100190, China*

[2]*Department of Physics and Astronomy, University of California, Irvine, CA92697, USA*

[3]*Institute of High Energy Physics, Chinese Academy of Sciences, Beijing 100049, China*

[4]*Shanghai Synchrotron Radiation Facilities, Shanghai Institute of Applied Physics, Chinese Academy of Sciences, Shanghai 201204, China*

[5]*Collaborative Innovation Center of Quantum Matter, Beijing 100190, China*

The origin of the non-trivial surface state of YbB$_6$, a topological insulator with mixed-valence rare-earth Yb, remains controversial. Contrary to the theoretical prediction, experiments suggest that the Yb-4$f$ electrons are not involved in the formation of the topological states at ambient pressure. High pressure is an effective way to change the valence states of the Yb ions and may help to understand the issue. Here, we report the first observation of two pressure-induced quantum phase transitions, from a topological insulator to an intermediate metallic phase and from the metallic phase to an exotic topological insulator respectively, in YbB$_6$. Our results reveal that the high-pressure insulating phase is associated with the pressure-induced valence change of the Yb $f$-orbitals and suggest that the two insulating states may be topologically different in nature as they originate from the *d-p* and *d-f* hybridizations, respectively. The tunable topological properties of YbB$_6$ revealed in this high pressure study may shed light on the intriguing correlation between the topology and the 4$f$ electrons. In this study, a comprehensive high-pressure methods are adopted, including resistance, Hall effect, X-ray diffraction and X-ray absorption measurements.

Topological insulators (TIs) have attracted intensive interest recently due to their non-trivial electronic properties of coexisting insulating bulk and topologically protected conducting surface [1-5]. Such exotic electronic states have been observed in the compounds of $Bi_2Se_3$, $Bi_2Te_3$ and HgTe [6-10]. A recent advance in the field is the discovery of a new type of TIs, namely the rare-earth hexaborides. As containing *f*-electrons in their configurations, these hexaboraides may exhibit exotic electronic correlation effects [11-14]. These TIs are expected to bridge the physics between heavy electron material and topological insulator and could host novel electronic states that are different from that of the usual TIs without *f*-electrons. $YbB_6$ and $SmB_6$ are the prototypical systems with such unusual properties, leading to the speculation of more exotic correlated topological surface state coexisting with the insulating bulk. Pressure is a powerful means to create new exotic electronic states by diverse ways, such as by tuning the band gap or changing the valence state of the rare-earth ions.

Experimentally, these non-trivial surface states have been identified by angle-resolved photoemission spectroscopy (ARPES) [15-18]. These studies demonstrate that $SmB_6$ is a strongly correlated topological Kondo insulator [19,20] with the intriguing feature of changing from a poor metal at room temperature to an insulator with residual conductance at low temperature [15-17,21,22]. High pressure studies on $SmB_6$ indicate that sufficient pressure can suppress its Kondo coherence and drive the system to an antiferromagnetic (AFM) metallic state [23,24]. $YbB_6$ crystallizes in the same crystal structure as $SmB_6$, but it presents very different electronic structures due to that Yb has a fully filled 4*f* shell in contrast to Sm's half-filled 4*f*-shell. Moreover, ARPES measurements on YbB$_6$ suggest that its topological surface state could only originate from the hybridization between Yb *d*-orbitals and Boron *p*-orbitals [25], in contradiction to the predicted *d-f* hybridization by theory [14]. To shed light on the nature of its TI state and explore potential pressure-induced new quantum phenomena, we performed resistance, Hall coefficient, X-ray diffraction and X-ray absorption measurements on pressurized YbB$_6$ in this study.

High quality single crystals of YbB$_6$ were grown by the Al flux method, as described in Ref. [26]. Pressure was generated by a diamond anvil cell (DAC) with two opposing anvils sitting on a Be-Cu supporting plate. Diamond anvils of 400 μm and 300 μm flats were used respectively for this study. Nonmagnetic rhenium gaskets with 200 μm and 100 μm diameter holes were used for different runs of the high-pressure studies. The four-probe method was applied on the (001) facet of single crystal YbB$_6$ for all high pressure transport measurements. To keep the sample in a quasi-hydrostatic pressure environment, NaCl powder was employed as the pressure medium for the high-pressure resistance, magnetoresistance and Hall effect measurements. High pressure X-ray diffraction (XRD) and X-ray absorption spectroscopy (XAS) experiments were performed at beam line 4W2 at the Beijing Synchrotron Radiation Facility and at beamline 14W at the Shanghai Synchrotron Radiation Facility, respectively. Diamonds with low birefringence were selected for the experiments. A monochromatic X-ray beam with a wavelength of 0.6199 Å was adopted for all XRD measurements. To maintain the sample in a hydrostatic pressure environment, silicon oil was used as pressure medium in the high-pressure XRD and XAS measurements. Pressure was determined by the ruby fluorescence method [27].

Figure 1 shows the results of high-pressure resistance measurements for a single crystalline sample of $YbB_6$. Although the sample exhibits metallic behavior in the temperature range 50-300 K at 0.9 GPa, at lower temperature an insulating behavior sets in, as manifested by a small upturn below 50 K (Fig.1a). This upturn is observed in three independent runs' measurements on samples obtained from different batches, consistent with its ambient-pressure behavior (Fig.S1 of Supplementary Information). Upon increasing pressure, the upturn is suppressed dramatically (Fig.1a) and eventually goes away at ~ 10 GPa (Fig. 1b). Subsequently, metallic behavior over entire temperature range in the resistance is observed in the pressure range 10.5-14.2 GPa (Fig.1b), reflecting the closure of the insulating gap in the bulk and the conversion from its non-trivial topological insulating state into a metallic state under pressure. Interestingly, an insulating behavior re-emerges at 15.2 GPa and becomes pronounced upon further increasing pressure (Fig.1b and 1c). At 19.8 GPa, an apparent resistance plateau is observed at very low temperatures, similar to what has been seen in the topological Kondo insulator $SmB_6$. A close inspection on the resistance-temperature curves finds a clear saturation of the resistance (inset of Fig.1c), which is one of the most prominent features for the metallic surface state of the topological Kondo insulator [26]. The onset temperature (T*) of the plateau varies with pressure and shifts to the higher temperatures upon increasing pressure (inset of

Fig.1c and Fig.1d). Since the lowest temperature of our instrument is 4 K, we cannot detect the T* (it should be lower than 4 K) for the sample subjected to pressure ranging from 15 GPa to 19 GPa. Extrapolation of the obtained T* down to lower pressure gives a critical pressure point at ~15 GPa (Fig.1d), where the insulating gap is closed. To determine the critical pressures for the insulator-metal-insulator transition, we plot the pressure dependence of the resistance at different temperatures (Fig.1e). It is seen that the resistance at all temperatures remains nearly unchanged between 10-15 GPa, but shows very different trends below 10 GPa and above 15 GPa, giving rise to two critical pressures: $P_{C1}$ (~10 GPa), which separates the non-trivial TI state at low pressure and the bulk metallic state, and $P_{C2}$ (~15 GPa), which separates the bulk metallic state and the bulk insulating state at high pressure.

To further characterize the pressure-induced changes in YbB$_6$, we performed high pressure Hall effect measurements. The Hall coefficient $R_H$ measured at 4 K is plotted as a function of pressure in Fig.1f. The $R_H$ displays negative value over the entire pressure range, revealing that electron carriers dominate in YbB$_6$. We find that below ~4 GPa the absolute value of $R_H$ decreases rapidly with increasing pressure. This remarkable decrease corresponds to the dramatic reduction of the resistance (Fig.1e). However, for pressures ranging from 10 GPa to 15 GPa, the $R_H$ barely changes, which may be attributed to the balance between electron and hole carrier population, suggesting that this metallic state might be a semi-metal [28]. Significantly, an increasing trend of change in the pressure dependence of $R_H$ is observed above 15 GPa. This increase in $R_H$ matches with the observed insulating behavior in the resistance (Fig.1b and 1c).

The pressure-induced insulator-metal-insulator transitions showing in the transport properties of YbB$_6$ are surprising because pressure usually increases the conduction/valence band width, makes the insulating gap overlapped and, as a result, drives the system into a metallic state. To clarify whether the physical origin of the gap closing and re-opening observed in the compressed YbB$_6$ is related to a structural phase transition, we conducted high pressure X-ray diffraction experiments for YbB$_6$ at beamline 4W2 at the Beijing Synchrotron Radiation Facility. We find that increasing pressure consistently pushes all Bragg peaks to a larger 2θ angle, but no new peak appears up to 31.2 GPa (Fig.2a). The lattice parameter exhibits linear pressure-dependence and the plot of pressure dependent volume shows no obvious discontinuity (Fig.2b and 2c). These results rule out the possibility of pressure-induced structural transition in YbB$_6$ and indicate that the closing and re-opening of the activation energy gap is driven by electronic phase transitions.

Figure 3a plots *dR/dT* curves measured in the temperature range 1.5-300 K for all pressurized samples. The activation energy ($\varepsilon_a$) as a function of pressure is estimated using the equation $R(T)=1/[R_S^{-1}+R_B^{-1}exp^{-\varepsilon a/2k_BT}]$, where the first component is the resistance from surface contribution and the second one represents the bulk resistance. We find a good fit to the experimental data in the insulating regime for all pressures (Fig. S2 of Supplementary Information). The solid dots in Fig.3a are the extracted activation energy ($\varepsilon_a$) of the sample subjected to different pressures. It is seen that, at the base temperature, the phase diagram contains three distinct regimes. In the left regime, $\varepsilon_a$ is reduced with increasing pressure and approaches zero at $P_{C1}$ (~ 10 GPa), demonstrating that the host sample undergoes a transition from a topologically non-trivial insulating (TI-1) state, with a positive *dR/dT* in the high temperature range and a negative *dR/dT* in the low temperature range, to a metallic (M) state (the middle regime). On further increasing pressure above $Pc_2$, the sample moves into another insulating state (the right regime).

Intriguingly, the low temperature resistance of YbB$_6$ at pressure above $Pc_2$ is very similar to that of SmB$_6$ at ambient pressure (Fig. 1c). It is hence natural to ask whether this high pressure insulating state is also topologically non-trivial, and if yes, whether it is a topological Kondo insulator. To address these issues, we applied a magnetic field of 7 Tesla (T) on the sample at given pressures and found an enhancement of its resistance in the whole temperature range for the pressure ranging from 16.3 GPa to 23.9 GPa (Fig.3b-d). The positive magnetoresistance effect found here excludes the existence of a strong Kondo effect in the pressurized YbB$_6$, because the Kondo effect is featured with a negative magnetoresistance [26]. However, the resistance plateau, observed clearly at lower temperature for the pressurized sample at 20.7 GPa and 24.9 GPa under zero field, vanishes under 7 T (inset of Fig.3c and Fig.3d). Similar resistance behaviors have been found in TKI SmB$_6$ and are believed to be the fingerprint of non-trivial metallic surface state [15,26]. This inspires us to propose that the new insulating state emerged at high pressure in YbB$_6$ may also be topologically non-trivial. Here, we denote the pressure-induced insulating state as (T)I-2.

We note that previous ARPES measurements and X-ray photoelectron experiments at ambient pressure have established that the divalent Yb component is dominant in the bulk YbB$_6$ while a trace amount of trivalent Yb component may exist only on the surface [17,25,29,30]. To clarify the underlying mechanism for the pressure-induced quantum phase transitions and the nature of the two topological insulating phases, we performed high-pressure X-ray absorption measurements at beamline 15U at Shanghai Synchrotron Radiation Facilities. Representative L$_{III}$-edge X-Ray absorption spectra of YbB$_6$ collected at different pressures are presented in Fig.4a. In the pressure range investigated, the Yb$^{2+}$ peak is clearly seen at 8939.8eV but no visible Yb$^{3+}$ peak is observed at higher energy. Instead, a dip is found at 8946.3eV where the Yb$^{3+}$ peak usually appears. Significantly, we find that the intensity of Yb$^{2+}$ peak decreases upon increasing pressure, while the dip is gradually filled up. A careful analysis on the extended X-ray absorption spectra for the sample indicates that the pressure-induced reduction of the dip should be attributed to the growth of the Yb$^{3+}$ component (Fig.S3 of Supplementary Information). As is shown in Fig. 4b, the absorption spectra collected at ambient pressure and 28.2 GPa clearly demonstrate how the pressure-induced enhancement of the Yb$^{3+}$ spectral weight compensates the observed dip. Taking the mean valence of Yb at ambient pressure as $v$=2, the pressure dependence of $v$ (Fig.4c) can be deduced from the relative intensities of the Yb$^{2+}$ and Yb$^{3+}$ components. We find that the $v$ remains nearly unchanged below 10 GPa, whereas it increases rapidly at pressure greater than 15 GPa where YbB$_6$ enters the (T)I-2 state.

Taking into account the corresponding increase of the $R_H$ and the gap re-opening (Fig.1f and Fig.3a), the above analysis suggests that the high pressure (T)I-2 state may be associated with the *d-f* hybridization, as has been theoretically predicted albeit for ambient pressure [14], in contrast to the low pressure TI-1 state that originates from the *d-p* hybridization, as has been shown in ARPES measurements [25]. The metallic phase in between may therefore involve complicated interplay of all three orbitals with their respective spin-orbit couplings [31] and valence changes. Finally, we propose that applying sufficiently high pressure may further increase the fraction of $Yb^{3+}$ and push the (T)I-2 state in $YbB_6$ to a Kondo-like TI state similar to that of $SmB_6$. Higher pressure studies on $YbB_6$ are needed to prove this, which is beyond the scope of the present study. Our high pressure experimental results exhibits a clean pathway connecting the usual topological band insulator and the exotic topological correlated insulator and may help to understand the discrepancy between theoretical prediction and experimental observation on $YbB_6$.

**Acknowledgements**

The work in China was supported by the NSF of China (Grant No. 91321207, 11427805, 11174339), 973 project (Grant No.2011CBA00100, 2015CB921303) and the Strategic Priority Research Program (B) of the Chinese Academy of Sciences (Grant No. XDB07020300, XDB07020200). The works in the USA were supported by DARPA under agreement number FA8650-13-1-7374 and FAPESP 2013/2018-0.

†To whom correspondence should be addressed.
E-mail: llsun@iphy.ac.cn, zhxzhao@ipny.ac.cn

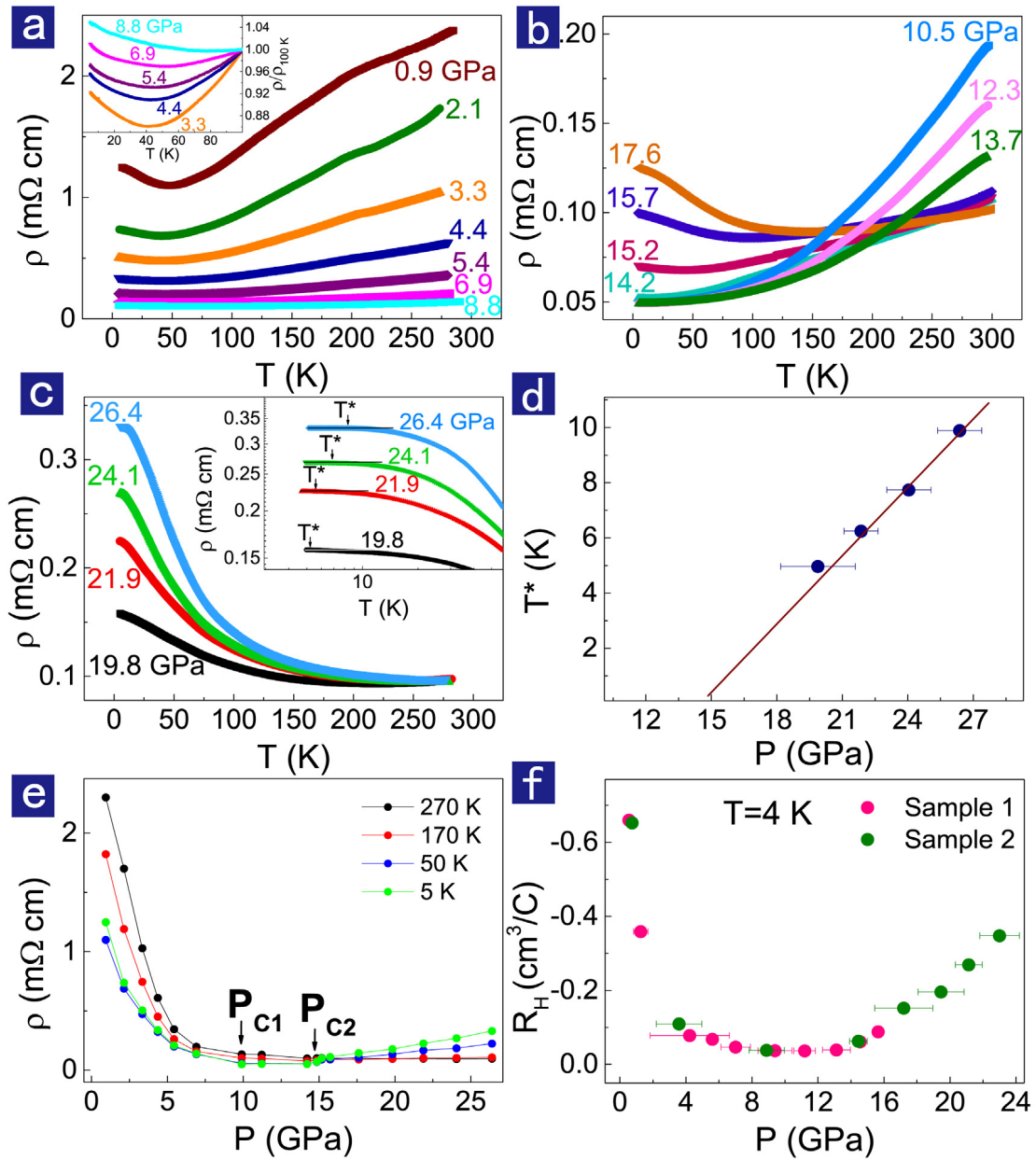

Figure 1 (a) Resistance-Temperature (*R-T*) curves measured in pressure range 0.9-8.8 GPa, showing a remarkable pressure-induced resistance decrease in full temperature regime. The inset displays normalized *R-T* curves in low temperature range, showing the upturns of the resistance measured at different pressures. (b) Selective *R-T* curves in the pressure range 10.5-17.6 GPa, displaying a dramatic change in transport. (c) The *R-T* curves measured in pressure range 19.8-26.4 GPa, exhibiting a significant pressure-induced enhancement of insulating behavior. The inset shows the residual resistance (T*) in lower temperature at different pressures. (d) T* as a function of pressure. (e) The pressure dependence of the resistance obtained at different fixed temperatures, illustrating two critical pressures ($P_{C1}$ and $P_{C2}$). $P_{C1}$ represents the transitions from the TI state to a metallic state and $P_{C2}$ represents the transition from the metallic state to another insulating state. (f) Hall coefficient ($R_H$) as a function of pressure for single crystal YbB$_6$ at 4K.

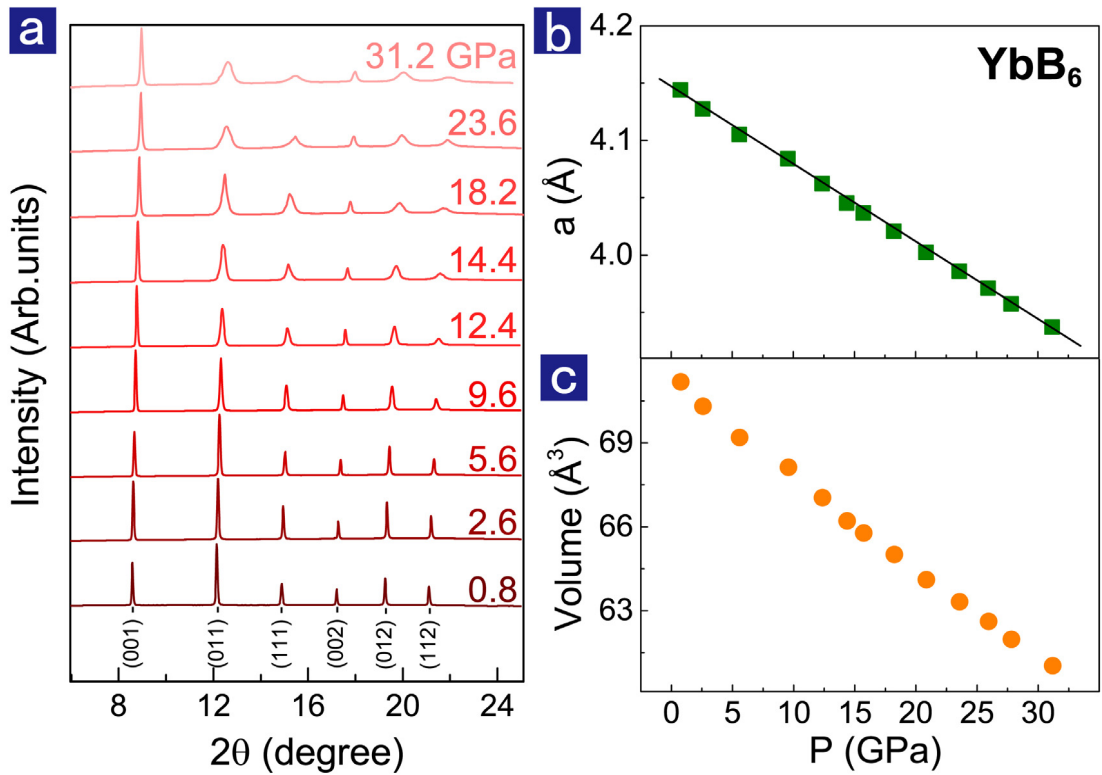

Figure 2 (a) X-ray diffraction patterns of YbB$_6$ collected at different pressures, showing that no structure phase transition occurs over entire pressure range investigated. (b) and (c) Pressure dependences of lattice parameter and volume of YbB$_6$.

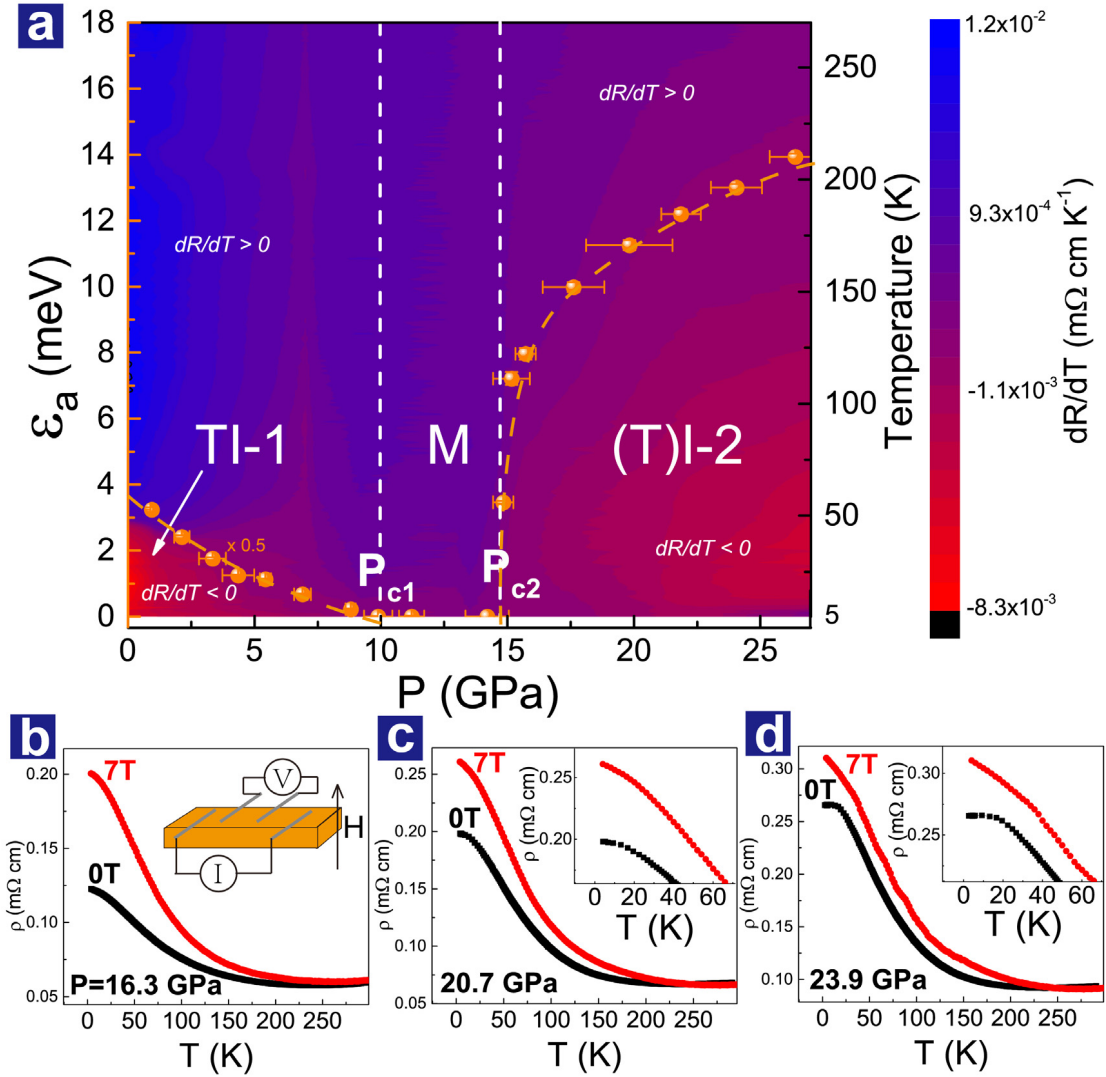

Figure 3 (a) Phase diagram of pressure dependence of activation energy gap (left axis) and temperature (right axis), as well as summary of *dR/dT* as a function of temperature and pressure. The solid dots represent the activation gap ($\varepsilon_a$). The value of the $\varepsilon_a$ in TI-1 regime is scaled by 0.5 to fit the pressure dependence of *dR/dT*. The acronyms TI-1 and (T)I-2 stand for the initial topological insulating state and pressure-induced insulating state, respectively. M represents metallic state. (b)-(d) Temperature dependence of resistance measured at zero field and 7 T at fixed pressure. The inset illustrates the details for the measurements. The acronyms, I, V and H, stand for current, voltage and magnetic field, respectively.

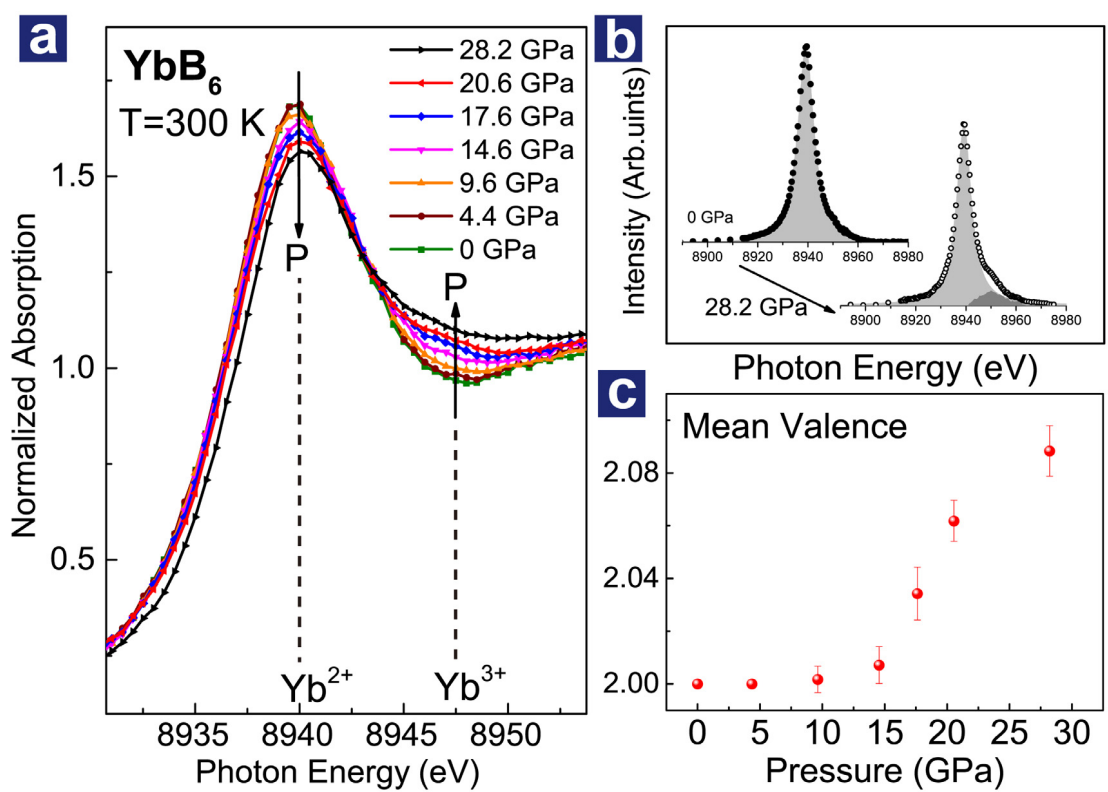

Figure 4 (a) X-ray absorption spectra (XAS) of Yb through the $L_{III}$-edge at different pressures. The dashed vertical lines indicate characteristic energy positions of $L_{III}$-edge peak for $Yb^{2+}$ and $Yb^{3+}$, respectively. (b) XAS of Yb measured at ambient pressure and 28.2 GPa, after the background deduction. The light gray area and dark gray area represent the $Yb^{2+}$ component and $Yb^{3+}$ component, respectively. (c) Pressure dependence of mean valence determined from XAS data.